\documentstyle[sprocl,epsfig,here]{article}

\newcommand{\tautau}{\mbox{$\tau^{+}\tau^{-}$}}
\newcommand{\ee}{\mbox{$\mathrm{e}^{+}\mathrm{e}^{-}$}}

\newcommand{\sqrts}{\mbox{$\sqrt {s}$}}
\newcommand{\bb}{\mbox{$\mathrm{b} \bar{\mathrm{b}}$}}
\newcommand{\mH}{\mbox{$m_{\mathrm{H}}$}}
\def\mA{\mbox{$m_{\mathrm A}$}}
\def\mh{\mbox{$m_{\mathrm h}$}}
\newcommand{\tanb}{\mbox{$\tan \beta$}}

\begin{document}
\begin{titlepage}
\def\thefootnote{\fnsymbol{footnote}}       

\begin{center}
\mbox{ } 

\vspace*{-3cm}

\end{center}
\begin{flushright}
\Large
\mbox{\hspace{10.2cm} hep-ph/9807566} \\
\mbox{\hspace{10.2cm} IEKP-KA/98-15} \\
\mbox{\hspace{11.7cm} July 1998}
\end{flushright}
\begin{center}
\vskip 0.5cm
{\Huge\bf
Aspects of 

Higgs Boson
\smallskip

Searches at LEP2}
\vskip 1cm
{\LARGE\bf Andr\'e Sopczak}\\
\smallskip
\Large University of Karlsruhe

\vskip 1.5cm
\centerline{\Large \bf Abstract}
\end{center}

\vskip 3cm
\hspace*{-3cm}
\begin{picture}(0.001,0.001)(0,0)
\put(,0){
\begin{minipage}{16cm}
\Large
\renewcommand{\baselinestretch} {1.2}
A preliminary combined mass limit of about 90~GeV has been reported 
for the Minimal Standard Model Higgs boson from the 183~GeV data 
taken in 1997, with a total luminosity of about 200~pb$^{-1}$. Similar 
mass limits have been reported from the individual LEP experiments 
based on the first 1998 data at 189~GeV with up to 40~pb$^{-1}$ each.
For the first time the LEP experiments have made a common effort 
to derive preliminary combined limits in the Minimal Supersymmetric 
extension of the Standard Model. For three MSSM benchmark parameter 
sets, mass limits for the lighter scalar of 77~GeV and for the 
pseudoscalar of 78~GeV have been derived. 
In a more general parameter scan unexcluded mass regions have been
revealed.
An independent variation of the SUSY parameters 
results in cancellation effects of production cross sections and 
thus some parameter regions are not excluded in this more general 
framework. Direct and indirect limits from other SUSY particle
searches have been taken into account.
\renewcommand{\baselinestretch} {1.}

\normalsize
\vspace{1cm}
\begin{center}
{\sl \large
Presented at the Sixth International Symposium on Particles, Strings, and
Cosmology, PASCOS--98, Boston, March 22--29, 1998
\vspace{-6cm}
}
\end{center}
\end{minipage}
}
\end{picture}
\vfill

\end{titlepage}


\newpage
\thispagestyle{empty}
\mbox{ }
\newpage
\setcounter{page}{1}

\title{ASPECTS OF HIGGS BOSON SEARCHES AT LEP2}

\author{ANDRE SOPCZAK}

\address{Karlsruhe University\\E-mail: andre.sopczak@cern.ch} 

\maketitle\abstracts{
A preliminary combined mass limit of about 90~GeV has been reported 
for the Minimal Standard Model Higgs boson from the 183~GeV data 
taken in 1997, with a total luminosity of about 200~pb$^{-1}$. Similar 
mass limits have been reported from the individual LEP experiments 
based on the first 1998 data at 189~GeV with up to 40~pb$^{-1}$ each.
For the first time the LEP experiments have made a common effort 
to derive preliminary combined limits in the Minimal Supersymmetric 
extension of the Standard Model. For three MSSM benchmark parameter 
sets, mass limits for the lighter scalar of 77~GeV and for the 
pseudoscalar of 78~GeV have been derived. 
In a more general parameter scan unexcluded mass regions have been
revealed.
An independent variation of the SUSY parameters 
results in cancellation effects of production cross sections and 
thus some parameter regions are not excluded in this more general 
framework. Direct and indirect limits from other SUSY particle
searches have been taken into account.}

\section{Introduction}
Over the last eight years the LEP experiments have made great
progress in their search for Higgs bosons. So far, despite 
immense efforts, no Higgs bosons have been found, and the quest to
understand mass generation continues. The very successful
operation of the LEP accelerator at the Z resonance until 1995 (LEP1) 
has continued with the successive energy increase (LEP2) to 
189 GeV in 1998, and to about 200 GeV over the next two years.
LEP2 operation has permitted a large sensitivity increase which has
so far manifested itself in larger mass limits.
In the Minimal Standard Model (MSM), mass generation results from
one complex Higgs doublet. The combined LEP1 mass limit of 
about 65 GeV\cite{gauhati} has been increased to the current limit 
of about 90 GeV\cite{lepwg98}.

This article is based on the results from the four LEP
experiments ALEPH, DELPHI, L3 and OPAL, and focuses on the 
combined limits and the interpretation in the Minimal
Supersymmetric extension of the Standard Model (MSSM).
The large interest in the search for Higgs bosons in the 
framework of the MSSM results from the fact that the MSSM
predicts a light Higgs boson 
which could be discovered at LEP as a first sign of Supersymmetry.

\section{Status of the Higgs Boson Searches}

\begin{figure}[t]
\caption[]{
Left: Preliminary MSM expected (dashed lines) and observed
(solid lines) confidence levels, $CL_s (m_{\rm H})$, 
obtained from combining the results of the four LEP
collaborations using the DELPHI statistical methods. 
The intersections of the curves with the 5\% horizontal 
lines define the 95\% CL lower bounds, 
expected and observed, for the mass of the SM Higgs boson.
Right: Preliminary MSSM benchmark exclusion plot for 
(\mh,\mA) at 95\% CL using the OPAL method to combine 
the different LEP experiments for $\tanb > 0.7$.
\label{fig:del98comb}}
\vspace*{-0.6cm}
\begin{center}
\epsfig{figure=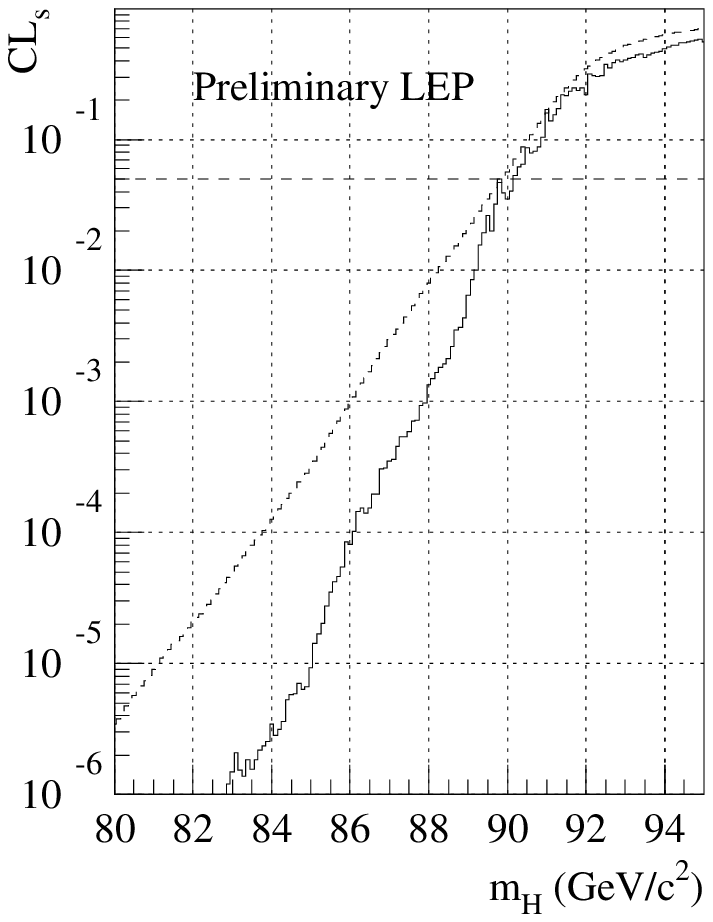,width=6cm}
\hspace*{-2cm} 
\epsfig{figure=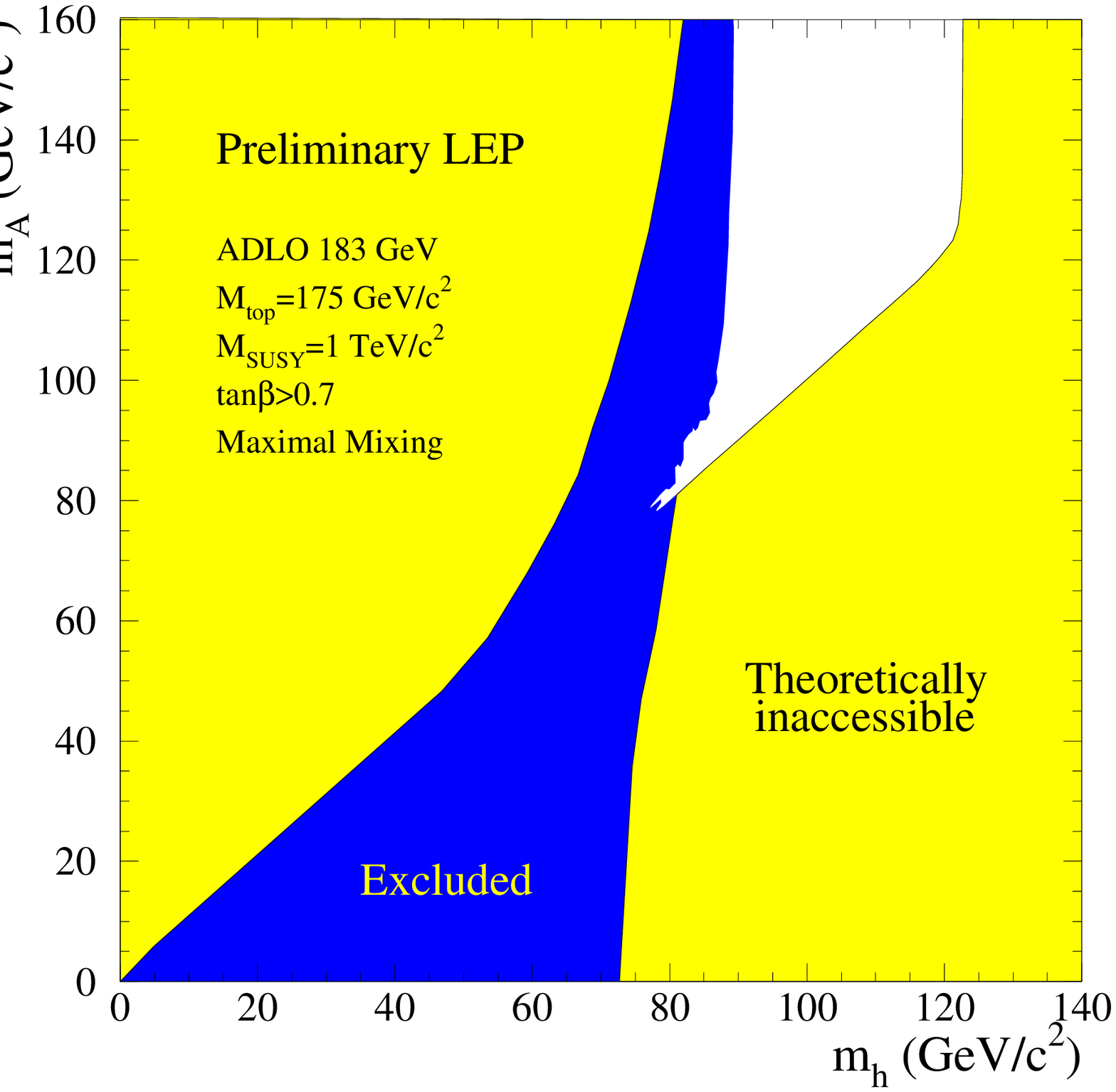,width=7.5cm} 
\end{center}
\vspace*{-0.5cm}
\end{figure}

The search for the Higgs bosons has been performed for many
different signatures. Each signature results from a possible
Higgs boson decay mode. For Higgs boson production,
the MSM process $\rm e^+e^- \rightarrow hZ$ gives a large
cross section if the Z is produced on-shell for sufficiently
large center-of-mass energy. In the MSSM, which predicts
five physical Higgs bosons, the production process
$\rm e^+e^- \rightarrow hA$ can also occur. The subsequent 
decay of the Z and Higgs bosons determines the search channel.
Most important are the Higgs boson decays into $\tautau$ and $\bb$.
Four conceptual different statistical methods give consistent 
limits when the results of the four LEP experiments
are combined\cite{lepwg97,lepwg98}.
The individual limits for each experiment are given in
Table~\ref{limit183} from ref.\cite{lepwg98}
The lowest combined mass limit of the four methods is 89.8~GeV, 
whilst the corresponding expected limit is 90.4~GeV. 
The spread of the observed limits from the four methods is only 
0.15~GeV. The confidence level (CL) curve using the DELPHI method is 
given in Fig.~\ref{fig:del98comb}.
Very preliminary results from the first 1998 data-taking with up to 
40~$\rm pb^{-1}$ per experiment already give similar mass 
limits\cite{treille} as for the total 1997 luminosity of about 
200~$\rm pb^{-1}$.
With the combined 1998 data, a Higgs boson of about 95~GeV could be
discovered, and by the end of the LEP2 programme in 2000, a Higgs boson
with a mass of up to 105~GeV could be found.

Figure~\ref{fig:del98comb} also gives an interpretation in the MSSM of the combined 
1997 data for a fixed set of SUSY parameters characterized by a large Supersymmetry
scale of 1~TeV, and maximal mixing in the scalar top sector with $A=\sqrt{6}$.
The exclusion confidence level has been calculated with the OPAL statistical
method and mass limits of about 77 and 78~GeV are derived for scalar and 
pseudoscalar Higgs bosons, respectively.

\begin{table} [hbtp]
\vspace*{-0.5cm}
\caption{Preliminary number of data and background events in the 183~GeV
data, as well as the expected and observed mass limits for the MSM Higgs 
boson at 95\% CL. 
Note that when deriving the ALEPH limit, no background subtraction
was applied. In the L3 analysis, the expected background and observed
number of events depend on the Higgs mass hypothesis; they are given here 
for \mH=85~GeV.
\label{limit183}}
\begin{center}
\vspace*{2mm}
\begin{tabular}{|l||c|c|c|c|}
\hline
Experiment                     & ALEPH & DELPHI & L3  & OPAL \\
\hline
Expected background            & 7.2  &    6.5  & 24.2 & 8.9  \\
Observed events                & 7    &     6   &  18  &   9  \\
\hline
Expected limit (GeV)           & 85.5 & 86.5    & 85.0 & 86.2 \\
Observed Limit (GeV)           & 87.9 & 85.7    & 87.6 &  88.3 \\
\hline
\end{tabular}
\vspace*{-0.5cm}
\end{center}
\end{table}

\section{MSSM Parameter Scan}
\vspace*{-0.2cm}
The importance of a parameter scan was pointed out for LEP1\cite{jras} 
and more detailed studies were performed when the LEP2 data
became available\cite{as97}.
New scan results from ALEPH\cite{alephscan}, 
DELPHI\cite{delphiscan} and OPAL\cite{opalscan} have
been released, and all parameter scans are based on a similar
variation of the SUSY parameters. 

\begin{figure}[htbp]
\vspace*{-2mm}
\caption{\label{fig:del183}
           Left: Preliminary DELPHI MSSM exclusion for 
           $\surd s = 183$~GeV and ${\cal L} = 54$~pb$^{-1}$.
           The region excluded by LEP1 (black), 
           the newly 95\% CL excluded region at LEP2 (grey),
           the region where the exclusion depends on the SUSY parameter
           set (white with dots), the region with no sensitivity (white),
           and the theoretically not allowed region (black) are shown.
           Right: Approximate combined LEP MSSM exclusion for 
           ${\cal L} = 200$~pb$^{-1}$.}
\vspace*{-3mm}
  \begin{center}
    \includegraphics[width=0.49\textwidth]{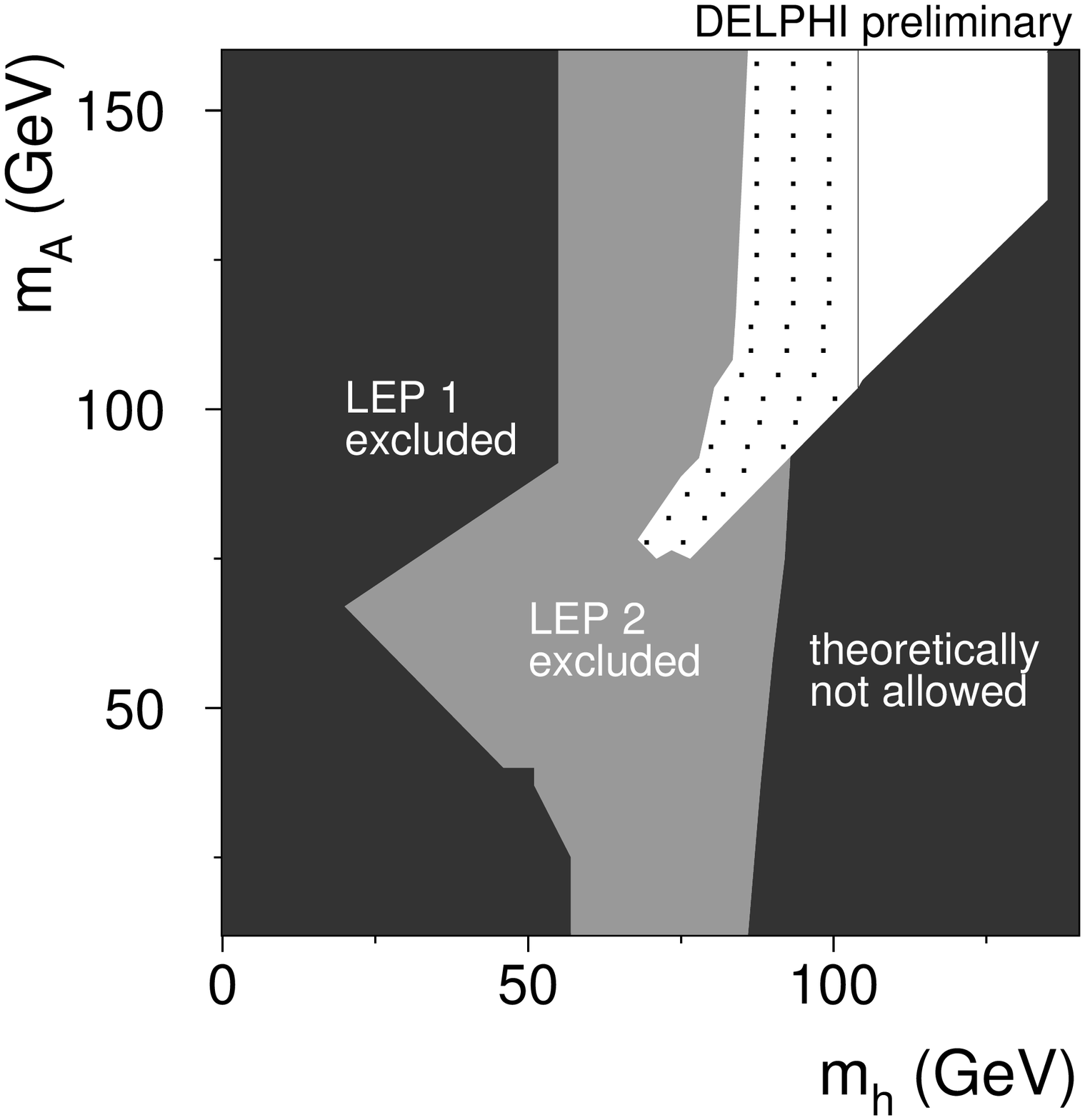}
    \includegraphics[width=0.49\textwidth]{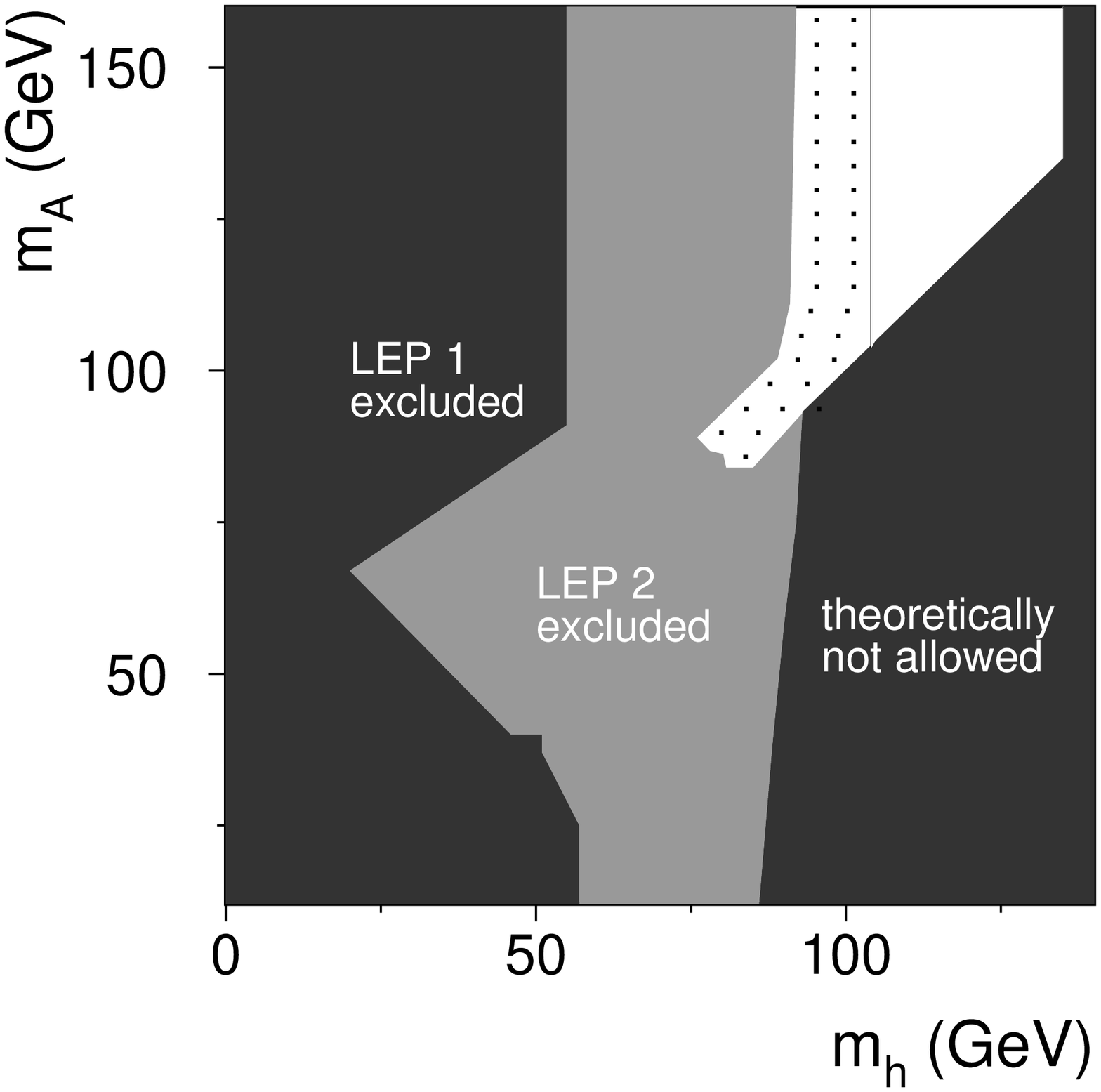}
  \end{center}
\vspace*{-0.9cm}
\end{figure}

Details of the DELPHI scan are given as an example. The DELPHI 
scan includes the Higgs and SUSY particle searches of the 183~GeV data.
A given (\mh,\mA) combination is excluded if, for all
SUSY parameter sets from the ranges defined in
Table~\ref{tab:allparameters}, the exclusion confidence level 
is larger than 95\% in at least one of the production reactions.
The parameters shown in Table~\ref{tab:allparameters} are the input
parameters for the calculation of the physical sfermion, chargino, and
neutralino masses. Some parameter combinations can be unphysical
(e.g. negative squark masses) or experimentally excluded.
Such cases are removed by imposing constraints
on neutralino and chargino particles\cite{delphi201}.
For each parameter combination of the scan, the exclusion confidence level
is calculated.
In addition, the limits from the reaction $\rm b \rightarrow s\gamma$ and 
the constraints on $\Delta\rho$ from light scalar top 
quarks\cite{chank} are taken into account. 
Figure~\ref{fig:del183} shows the excluded regions in the (\mh,\mA) 
plane for the preliminary DELPHI results\cite{delphiscan} and for
an approximate combination of all four experiments based on ref.\cite{as97}
OPAL obtains similar results with their parameter scan\cite{opalscan}.
The ALEPH parameter scan\cite{alephscan} points out that only a small 
fraction of the parameter combinations in the MSSM is not excluded 
compared to the benchmark results.

\begin{table}[htb]
\vspace*{-3mm}
\caption{\label{tab:allparameters}
Ranges of SUSY parameters used for independent variation in the
study of the MSSM neutral Higgs boson searches.}
\vspace*{-3mm}
\begin{center}
\begin{tabular}{|c|c|c|c|c|c|}
\hline
Parameter& $m_{\mathrm{sq}}$ (GeV) & $m_{\mathrm{g}}$  (GeV) &
$\mu$   (GeV) & $A$ & $\tan\beta$ \\
\hline
Range    & 200\,---\,1000   &  200\,---\,1000   & $-500$\,---\,500 &
$-1$\,---\,$+1$ & $0.5$\,---\,$50$ \\
\hline
\end{tabular}
\end{center}
\end{table}

The DELPHI scan over the SUSY parameter space gives a limit of about 67~GeV
on the scalar Higgs boson mass\cite{delphiscan} compared with a
limit of 74.4~GeV for the three sets of SUSY benchmark parameter 
sets\cite{delphi200}. 
The existence of newly unexcluded regions can be understood in the 
following way:
The unexcluded region begins near the $\mh+\mA=\sqrt{s}$ limit 
for $\mh > 67$~GeV.
In this range, the bremsstrahlung cross section $\rm e^+e^- \rightarrow hZ$  
can be small for some SUSY parameters. 
The complementary process $\rm e^+e^- \rightarrow hA$ 
is reduced kinematically, and thus no signal can be observed.

The 1997 combined LEP data of about 200~$\rm pb^{-1}$ leads to mass limits 
of about 76 and 83 GeV for scalar and pseudoscalar Higgs bosons, respectively.
Despite the different scan method and the additional direct and indirect
limits from other SUSY particle searches applied in Fig.~\ref{fig:del183},
very similar mass limits are obtained compared to the benchmark limits
in Fig.~\ref{fig:del98comb}. As pointed out in ref.~\cite{as97}, the combined
luminosity of the four LEP experiments is important for a larger coverage 
of the MSSM parameter space. In the example of Fig.~\ref{fig:del183}, the
limit on the scalar Higgs boson mass increased by 9~GeV.

\newpage

\section*{Acknowledgments}
\vspace*{-1mm}
I would like to thank the organizers of the conference 
for their kind hospitality, and Wim de Boer for his advice 
about the manuscript.

\end{document}